\documentclass[useAMS,usenatbib]{mn2e}
\usepackage{graphicx}
\usepackage{epsfig}
\newcommand{\LA}{\mbox{\raisebox{-0.6ex}{$\stackrel{\textstyle<}{\sim}$}}}
\newcommand{\GA}{\mbox{\raisebox{-0.6ex}{$\stackrel{\textstyle>}{\sim}$}}}
\begin{document}
\title{Models of the Compact Jet in GRS 1915+105}
\author[Brian Punsly] {Brian Punsly \\ 4014 Emerald Street No.116, Torrance CA, USA 90503 \\
ICRANet, Piazza della Repubblica 10 Pescara 65100, Italy,\\
E-mail: brian.punsly@verizon.net or brian.punsly@comdev-usa.com\\}

\maketitle \label{firstpage}
\begin{abstract} In this article, models are constructed of the compact jet in GRS 1915+105
during an epoch of optimal data capture. On April 02, 2003, the
object was observed in the hard X-ray/soft gamma ray band
(INTEGRAL), hard X-ray band (RXTE), near IR (ESO/New Technology
Telescope) and the VLBA (8.3 GHz and 15 GHz). The source was in a
so-called "high plateau state." The large radio flux provides high
signal to noise ratios in the radio images. Thus, one can image the
jet out to large distances ($ > 10^{15}$ cm). This combined with the broadband
coverage make this epoch the best suited for modeling the jet. The
parametric method developed in the papers
\cite{ghi85,ghi89,ghi96,sam97} that has been successfully utilized
in the realm of extragalactic radio jets is implemented. The basic
model is one where external inverse Compton (EIC) scattering of
accretion disk photons by jet plasma provides the hard X-ray
powerlaw. Unlike AGN jets, it is found that the radio jet must be
highly stratified in the transverse direction in order to produce
the observed surface brightness distribution in the radio images.
Various jet models are considered. The jet power is $Q \approx 3-4
\times 10^{38}$ ergs/sec if the hard X-ray powerlaw luminosity is
from EIC in the jet and $Q \approx 2 - 9 \times 10^{37}$ ergs/sec if
the X-rays are emitted from the accretion disk corona. These estimates
indicate that the jet power can be as high as 60\% of the total X-ray luminosity.

\end{abstract}
Galactic black hole accretion systems might serve as laboratories
for AGN (active galactic nuclei) on highly contracted time scales.
The black hole candidate, GRS 1915+105, routinely ejects powerful
radio emitting plasma out to large distances from the central black
hole at relativistic speeds and might be the Galactic analog of
radio loud AGN \citep{rod99,fen99,rus10}. These major flares are relatively rare compared to
the so-called compact jet state \citep{rus10}. A compact, optically thick,
elongated structure is frequently observed with long baseline
interferometry \citep{dha00,fuc03,rib04}. The low frequency emission is significantly less
than in a major flare. It is not clear if there is an analogous
structure in the context of AGN. The total power of the jet is also
unclear. In this article, the analogy with AGN is pursued by modeling
the jet using the standard AGN technique of \cite{ghi85,ghi89} with
the addition of external Compton scattering (EIC) from
\citep{ghi96,sam97}. There is a correlation or more precisely a
direct association of the formation of a compact jet with an excess
of hard X-ray emission in the form of a power law \citep{rus11,rod08}. The
primary assumption of the models of the jet presented below is that
this X-ray emission is produced in the base of the jet
\citep{mar05}. A scenario in which the X-rays are from Compton scattering in the accretion
disk corona is also explored briefly in this article with the same techniques.
\par Since most of the luminosity is in the X-ray band, complete high energy coverage is mandatory
for jet modeling. Furthermore, one needs the near IR photometry in
multiple bands in order to constrain the synchrotron spectrum. Also,
to break the degeneracy in the available models, one requires a
faithful representation of the surface brightness distribution of
the jet. It is a much simpler task to fit the jet spectrum if the
distribution of emissivity is ignored. To meet this challenge in the
presence of intrinsic source variability, one needs simultaneous
full coverage of the X-ray emission to high frequencies, IR
photometry and sensitive VLBA imaging. The only simultaneous set of
observations capable of fulfilling these requirements was the
campaign presented in \citep{fuc03,fuc04}. The strongest compact jet
ever imaged with VLBA was measured simultaneously in the near IR
(ESO/New Technology Telescope) and throughout the hard X-ray and
soft gamma ray band (INTEGRAL and RXTE). For more details on the
observations see \cite{fuc03,fuc04}.
\begin{figure}
\includegraphics[scale=0.35]{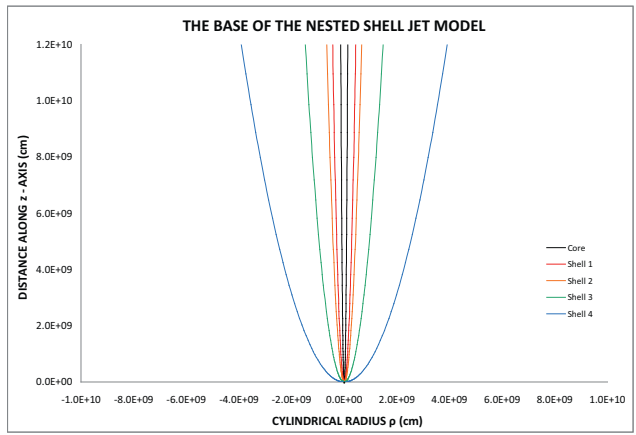}
\caption{The shell model of transverse jet structure.}
\end{figure}

\section{Some Preliminaries on Jet Speed} There is uncertainty in
the bulk velocity of the compact jet. The Doppler enhancement of the
intrinsic radiation is a critical parameter in the determination of
the jet power, $Q$. The Doppler factor, $\delta$, is given in terms
of $\Gamma$, the Lorentz factor of the outflow; $\beta$, the three
velocity of the outflow and the angle of propagation to the line of
sight, $\theta$; $\delta=1/[\Gamma(1-\beta\cos{\theta})]$
\citep{lin85}. For a jet, the apparent flux density, $S_{\nu}$ is
related to the intrinsic value by $S_{\nu,\, \mathrm{obs}}=
\delta^{2 + \alpha}S_{\nu,\,\mathrm{int}}$, where $\alpha$ is the
spectral index (the convention adopted in this paper is $S_nu \propto nu^{-alpha}$). Assuming an intrinsically symmetric bipolar jet, the
elevated flux of the approaching jet has been used to estimate
$\delta$. Using a value of $66^{\circ} < \theta< 70^{\circ}$ that
was deduced from discrete relativistic ejecta at other epochs (see
\cite{mir94,fen99}), estimates of $0.1 < \beta <0.54$ for the
compact jet at various frequencies and epochs of VLBI observations
have been obtained \citep{dha00,rib04}. For the epoch April 02,
2003, \cite{rib04} found disagreement between the jet velocity
inferred from the flux asymmetries of the 15 GHz and 8.3 GHz VLBA
images. They reconciled this by looking at the Clean Components,
concluding that a value of $\beta= 0.38 \pm 0.04$ was compatible
with the data. Thus, $\beta = 0.4$ is the baseline value in the
following calculations.
\par The discussion above is based on direct application of the data to
relativistic beaming as was done for discrete ejecta in major flares \citep{mir94,fen99}.
In \citet{dha00}, the difference in the directly inferred speeds of the compact jet
and the major flares was disconcerting. Thus, they conjectured on other explanations
including undetected relativistic motion. Thus, before the \citet{rib04} estimated value was adopted
in the jet models, an open mind was kept in regards to the value
of $\beta$. Preliminarily, a relativistic jet was considered as
suggested in \cite{dha00}. Assuming $\beta=0.98$ and
$\theta=66^{\circ}$ as was deduced from the kinematics of the
discrete ejections in \cite{fen99}, it is found that the flux of
blackbody radiation from the accretion disk is highly redshifted in
the frame of reference of the jet plasma, so synchrotron self
Compton emission (SSC) dominates the external Compton emission (EIC)
and SSC is the source of the power law X-rays. Many models were
constructed, but because the X-ray luminosity (of the unresolved
base of the bipolar jet) observed at earth is redshifted by a factor
of $\delta^{4}$, the intrinsic X-ray luminosity
$L_{\mathrm{int}}(\mathrm{X-ray}) \approx 3.5 \times
10^{40}\mathrm{erg/sec}$ \citep{lig75}. Therefore the jet power, $Q
> 3.5 \times 10^{40}\mathrm{erg/sec}$. This seems unreasonably
large. Consequently, relativistic bulk motion is not considered in
the following. These calculations indicate that a reasonable model
should have at most mildly relativistic velocities that are
consistent with the observed jet asymmetry (i.e., $\beta \approx
0.4$). EIC is therefore the source of the power law X-ray hard
tail (not SSC) since the redshifted thermal X-ray flux is much
larger than the local synchrotron flux within the base of the jet.
This result is considered in the context of other estimates of the jet velocity in the literature
in the Discussion section.
\begin{figure*}
\includegraphics[scale=0.35]{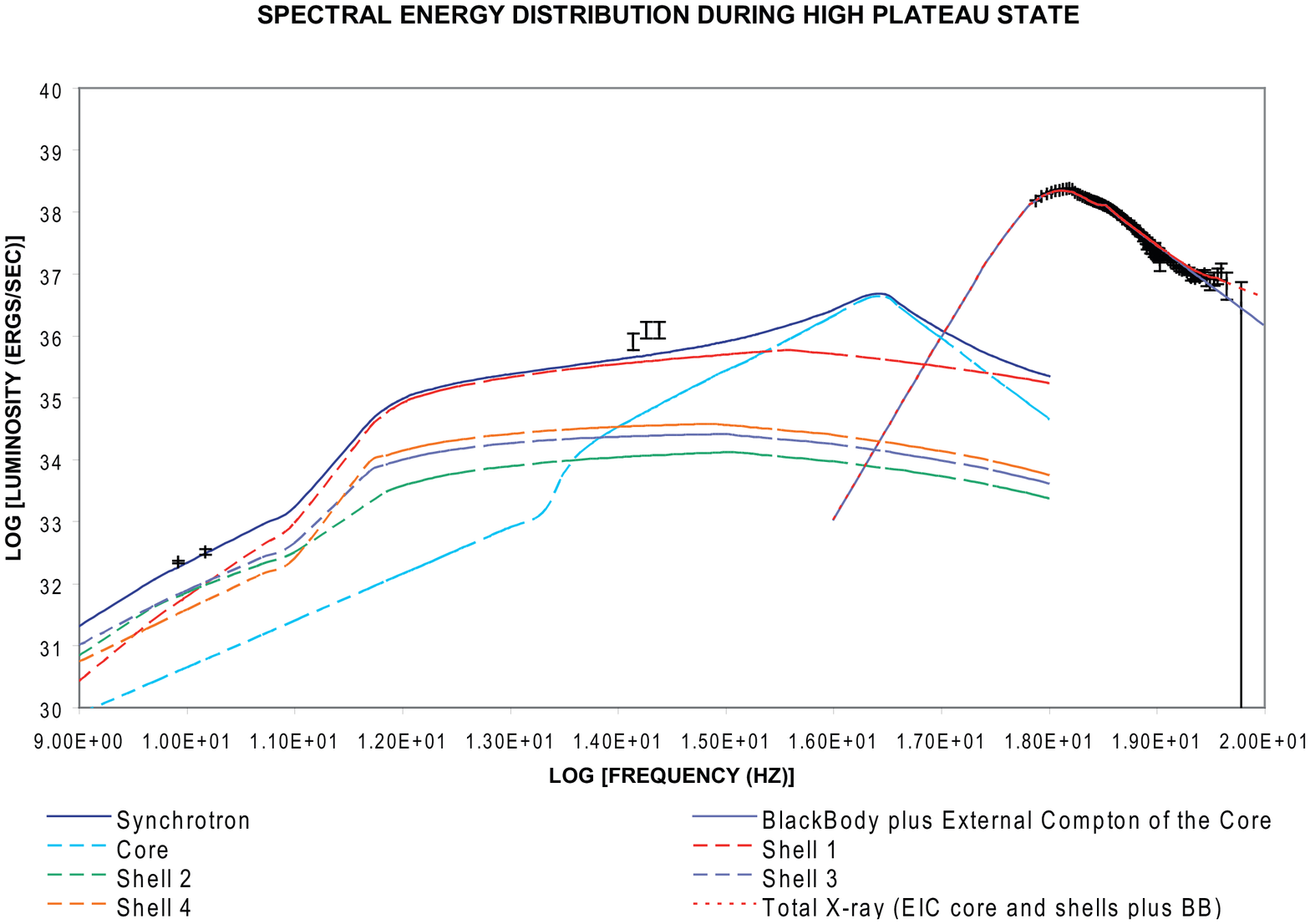}
\includegraphics[scale=0.35]{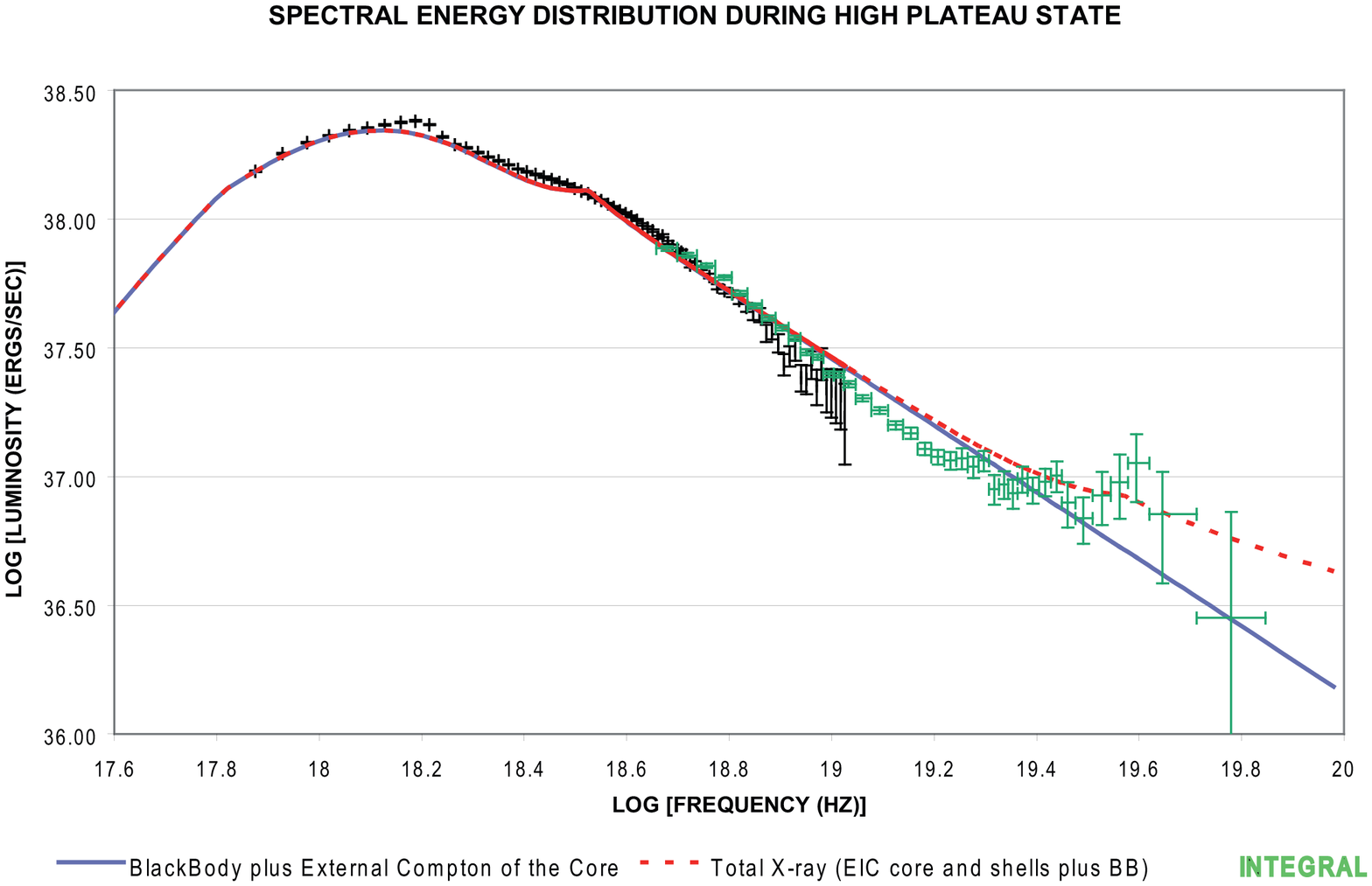}
\caption{The model fit to the broadband SED is in the left frame.
The IR error bars represent the variability at the time of
observation from Fuchs et al (2003). The radio data is from a
simultaneous VLA measurement at 8.3 GHz and a 15 GHz Ryle telescope
observation 2.5 hours earlier. The error bars represent the
variability seen in the radio measurements over a two hour time
scale (the length of the VLBA observation). This plot includes
contributions from the oppositely directed, but otherwise identical
counter-jet. The model has an excess above a simple power law in the
sub-mm as in the epochs observed in Ogley et al (2000). The right
frame is a close-up of the spectral energy distribution in the X-ray
region. The INTEGRAL data is in green and the RXTE data is in black.
The blue curve is the disk blackbody spectrum plus the EIC steep
powerlaw from the core. The red curve also includes the EIC from
shells 1 to 4}
\end{figure*}
\section{Description of the Jet Model and the Fit to the Data} In this section, the parameters in the jet
model are defined. One should review \cite{ghi85,ghi89} for further
discussion. The jet is parameterized in cylindrical coordinates,
$(\rho,\, \phi,\, z)$. The jets models are axisymmetric and have no
transverse gradients (this assumption is modified below). The jet shape and
axial gradients are parameterized by, $z_{0}$, $z_{\mathrm{max}}$,
$a$, $\epsilon$, $m$, $n$, $\Gamma$, $\gamma_{\mathrm{min}}$ and
$\gamma_{\mathrm{max}}$ through the the expressions,
\begin{eqnarray}
&&\rho = az^{\epsilon}; \, z_{0}< z < z_{\mathrm{max}}, \;
N(\gamma)=K\gamma^{-p}, \;\nonumber \\
&& K = K_{0}(z/z_{0})^{m\epsilon},\; B =
B_{0}(z/z_{0})^{n\epsilon}\; .
\end{eqnarray}
In equation (1), the electron energy is $E = \gamma m_{e}c^{2}$ and
the particle number density is $N = \int{N(\gamma)d\, \gamma}$. The
magnetic field strength is $B$. The jet begins and ends at $z_{0}$
and $z_{\mathrm{max}}$, respectively. The models typically have two
regions. The base of the jet is called the paraboloid, which is the
stronger inverse Compton source, and it is characterized in
\cite{ghi85} by $\epsilon=0.5$. The outer part of the jet is called
the conical region typically with $\epsilon=1$. This is generalized
to other values of $\epsilon$ for the outer jet in the following.
The outer jet is the primary source of the radio emission. Tables 1
and 2 list the jet model parameters and physical quantities,
respectively, that characterize a jet solution. The dimensional input parameters
and power law indices in Table 1 define the model in terms of equation 1. The
chosen value is either fitted to the data or set by the dimensions of the physical system
as noted in the table. All the quantities
in Table 2 are physical characteristics of the jet model. It is noted in the table that the
values are either an input parameter fitted to the data or a derived
physical quantity that is computed from the model. The jet plasma
is protonic. Notice that Tables 1 and 2 list four shells and a core
(shell 0) instead of one jet. The reason for this necessity was the
surface brightness distribution in the VLBA images is very extended.
A jet with no transverse gradients that accurately models the SED
always produces a surface brightness distribution that decays too
rapidly with $z$ to represent the VLBA maps (see Section 4). Figure 1
shows the relative scale of the nested shells near the base.  A
continuous transverse gradient would provide a smoother solution.
However, the simple crude model of 5 uniform (there are no
transverse gradients within each zone) zones is shown to be a
reasonably accurate representation of the spectrum and the
brightness profile in the following.
\par Each shell is parameterized in the tables.
Columns 2, 3, 4, 10 and 11 of Table 1 define the parameters for the
base of the jet in terms of equation (1) for each shell. Similarly,
columns 5 - 9 are the outer jet parameters. The initial data surface
from which the jet solution begins is formally defined for each shell
"i" as the set of points $\rho_{\mathrm{base}}(i-1)< \rho < \rho_{\mathrm{base}}(i)$
such that $z= z_{0}(\mathrm{base})$, where $z_{0}(\mathrm{base})$ is
the minimum axial coordinate in shell "i" and $\rho_{\mathrm{base}}(i-1)=0$
for $i =0 $. One important modification to the procedure of
\cite{ghi85} is that the inner shells must transfer radiation not
only through the optical depth provided by its own synchrotron self
absorption, but also that of the surrounding shells. This affects
the turnover frequency between the optically thick region of the
spectrum and the optically thin region of the spectrum through
equation (11) of \cite{ghi85}. Because the opacity dies off with
cylindrical radius, this is a second order correction to the
spectrum from each shell, but is retained to improve the accuracy of
the calculations. In the following models, $\theta = 70^{\circ}$ is
assumed as was determined to be consistent with the kinematics of
discrete relativistic ejections in \cite{mir94}. The flux densities
are converted to luminosity using a distance of 12.5 kpc to the
source as estimated in \cite{mir94}.
\par The baseline fit to the broadband spectral
energy distribution (SED) is in Figure 2 with the X-ray portion
magnified in the right hand frame. These plots are the analog of
Figure 5 from \cite{fuc03}. The X-ray data were calibrated and
reduced with current software suite, HEASOFT V 6.10 and OSA V 9.0,
that are improvements from \cite{fuc03,fuc04}. The new results were
generously provided by Jerome Rodriguez. The fit in the X-ray region
is not that good around $10^{19}$ Hz. There is increased curvature
between $6 \times 10^{18}$ Hz and $10^{19}$ Hz that is more evident
in the new reduction of the data than in \cite{fuc03}. There is an
overlap of the data in the HEXTE and the INTEGRAL spectrum in this
region as discussed in \cite{fuc03}. Even though the two data sets
do not agree in this area (the INTEGRAL spectrum is flatter with higher flux levels, see the related discussion in \citet{fuc03}), 
this does not appear to be the origin of
the spectral curvature, but coincidental. This region is not well
represented by the EIC powerlaw from the jet core (shell 0) in blue
and is actually exacerbated by the spectral hardening that seems to
be real at frequencies above $10^{19}$ Hz. In the model, the hard
excess is fit by the red, total luminosity curve which includes a
second EIC contribution from the surrounding shells. In this
decomposition, the spectral curvature at frequencies just below
$10^{19}$ Hz is consistent with spectral ageing from Compton losses
of the higher energy electrons in the core jet which have the
shortest radiative time scale, $\tau \sim \gamma^{-2}$
\citep{tuc75}. This suggests that a more elaborate jet model is
indicated for the core. In \cite{rod04,rod08}, the poor powerlaw fit
near $10^{19}$ Hz is improved by including an inverse Compton coronal
component with a thermal cutoff. The extra parameters introduced by
a corona with cutoff is beyond the scope of this paper, which is an
exploration of how well a simple jet model can explain the data.
\par Figure 3 is the surface brightness distribution based on the
modeled jet emissivity as it would appear with the resolution
provided by the synthetic beamwidth of the VLBA as modified by the
wavelength dependent interstellar scattering ($\Delta \theta \approx
0.15 (\lambda /1 \mathrm{cm})^{2}$ mas) that was determined in
\cite{dha00}. The plasma in the outer shells produce the low
frequency emission from innermost part of the jet, "the radio core" (or peak flux per beamwidth)
and the observed base of the radio jet (the partially resolved emission located within one
beammwidth of the core, 33.6 AU at 8.3 GHz and 17.0 AU at 15 GHz) that
is depicted by the brightness distribution in Figure 3. Shell 1
produces the farthest reaches of the jet. The lowest contour is at 15
times the RMS noise of the radio images, thus the vertical error
bars are very small on all the data points. The dominant source of error is the sparse u-v
coverage (the missing 8.3 GHz flux compared to the VLA measurement noted in the figure caption). There is really no way to attach
these errors to any of the individual contours. All that one can say is that the relatively long baselines of the VLBA favors missed flux on the least compact scales, the outer-most contours.

\begin{figure*}
\includegraphics[scale=0.35]{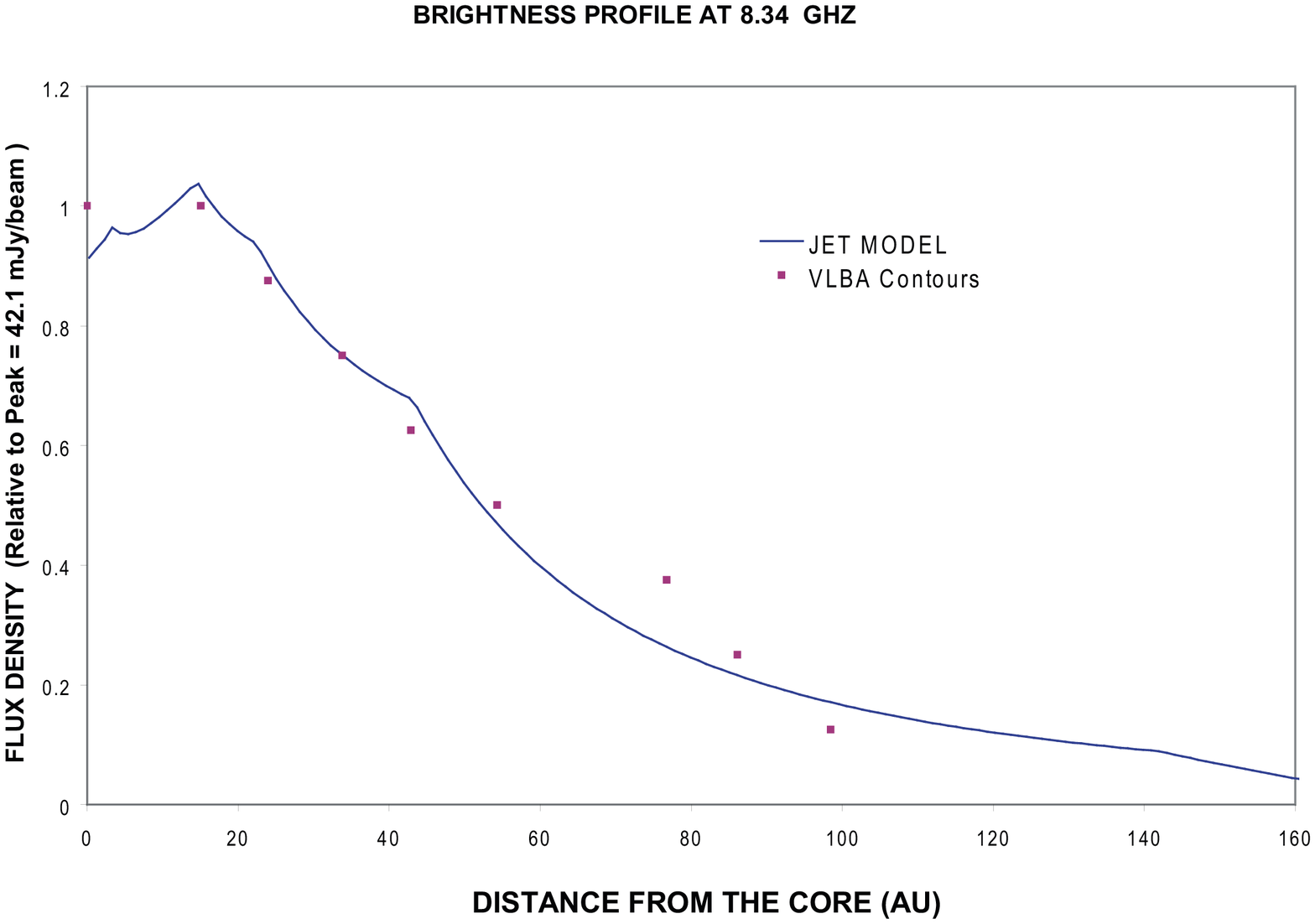}
\includegraphics[scale=0.35]{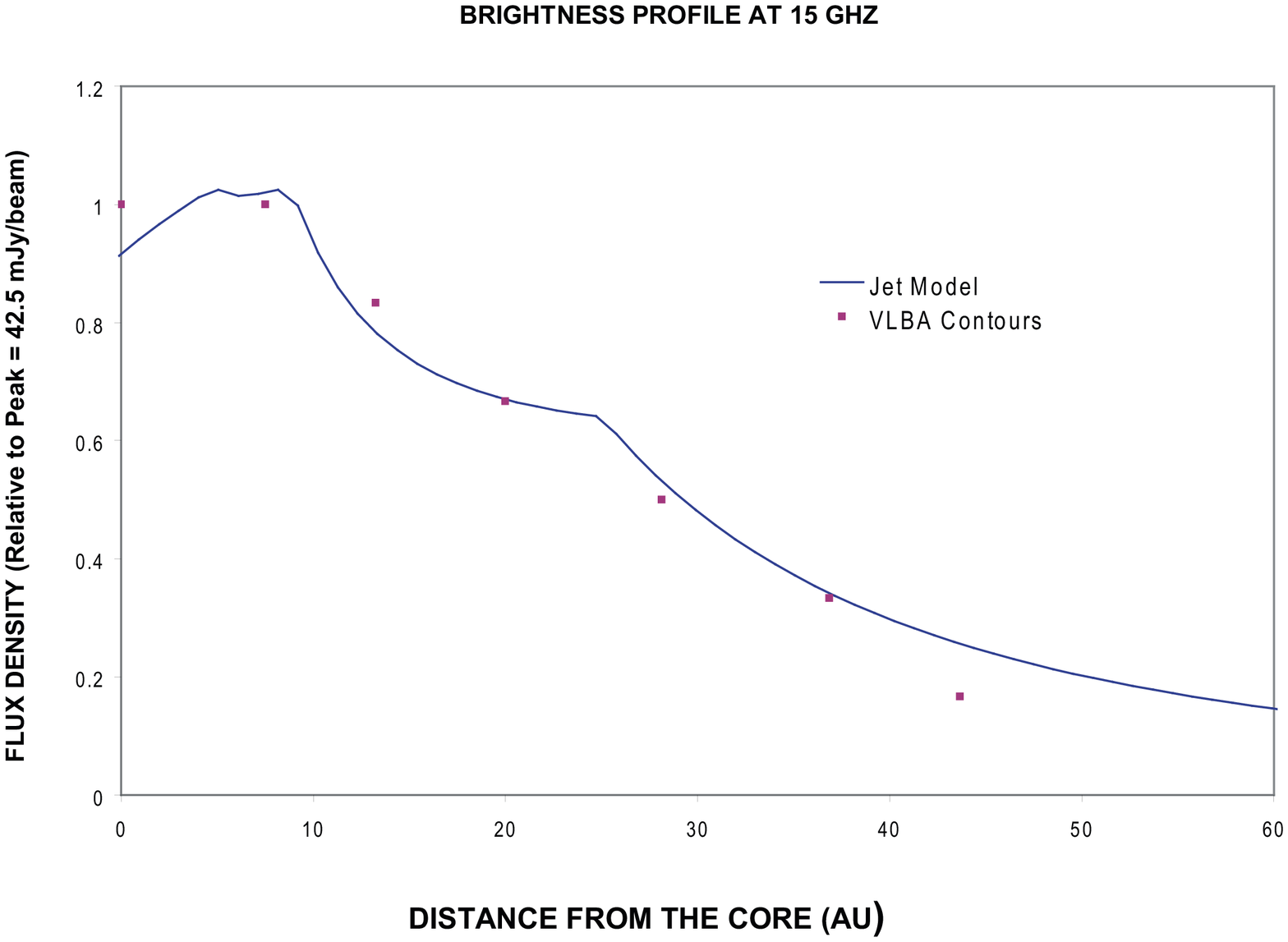}
\caption{The brightness profiles at 8.3 GHz and 15 GHz obtained from
the contours of the VLBA images in Fuchs et al (2004). Each dot
represents a contour level. For comparison, the synthesized flux per
beam that is produced from the 5 shell jet model convolved with the
VLBA beamwidth and interstellar scattering is plotted. The diffuse
nature of the jet emissivity requires a very broad base and
transverse gradients in the jet, greatly increasing the volume and
the required power. The excess flux at large $z$ provided by the
model is a desired effect, since simultaneous VLA measurements at
8.3 GHz in Fuchs et al (2003) show that $>$ 20\% of the flux is
resolved out by VLBA and likely resides at large distances, $> 10^{15}$ cm. The
plots do not cover the counter-jet as it was noted in Ribo et al
(2004) and in the text that a single velocity cannot produce the
asymmetry in the velocity profile, so there cannot be a good fit to
this data simultaneously. This might be evidence that transverse and
axial gradients in the bulk velocity are required or the jet model
is not representative of the emission.}
\end{figure*}
\begin{table*}
\caption{Input Parameters of the Jet Component Models}
{\footnotesize\begin{tabular}{ccccccccccc} \hline
Shell &  $m_{\mathrm{base}}$ & $n_{\mathrm{base}}$ & $a_{\mathrm{base}}$ & $m_{\mathrm{outer}}$  & $n_{\mathrm{outer}}$ & $a_{\mathrm{outer}}$& $\epsilon$ & $z_{0}(\mathrm{outer})$ & $z_{0}(\mathrm{base})$ & $\rho_{base}$ \\
 &   &  &  &   &  & &  & cm & cm & cm   \\
 & a   & b  & a   & a   & b  & a  & a  & a  & a  & c   \\

\hline
0   &  2.64  & 2 & $1.52 \times 10^{3}$  & 2.4 & 2 & $3.82 \times 10^{3}$ & 0.45 & $1.0\times 10^{8}$  & $5.0\times 10^{6}$  & $3.4\times 10^{6}$  \\
1   &  1.08  & 2 & $4.02 \times 10^{3}$ & 2.1 & 2 & $2.20 \times 10^{2}$ & 0.595 & $1.5 \times 10^{13}$  & $5.0\times 10^{6}$  & $8.8\times 10^{6}$  \\
2   &  1.02  & 2 & $6.03 \times 10^{3}$ & 1  & 2 &  1 & 0.8 & $4.0 \times 10^{12}$   & $5.0\times 10^{6}$  & $1.4\times 10^{7}$  \\
3   &  0.99  & 2 & $1.35 \times 10^{4}$ & 1 & 2 &  $5.0 \times 10^{-2}$ & 0.92 & $8.5 \times 10^{12}$   & $5.0\times 10^{6}$  & $3.0\times 10^{7}$   \\
4   &  1.00  & 2 & $3.57 \times 10^{4}$ & 1 & 2 & $1.4 \times 10^{-2}$ & 1 & $6.5 \times 10^{12}$   & $5.0\times 10^{6}$  & $8.0\times 10^{7}$  \\
\end{tabular}}
\caption {Table notes: a. input parameter for model fitted to SED
data, b. set by baryon conservation, c. input parameter is a minimum value consistent with the
black hole dimensions}
\end{table*}

\section{Physical Interpretation of the Jet Model}
The model presented above and described by the input parameters in Table 1
and Table 2 can be interpreted physically in terms of quantities like mass
flux, Poynting flux and magnetic dominance. In the following, it
is described how the model parameters relate to physical quantities in
Table 2. Column 2 of Table 2 gives the energy flux in each shell at
the base. The next 4 columns provide the number density, magnetic
field morphology and component strength. Columns 7 and 8 are the ratio of magnetic energy flux to mechanical
energy flux at the minimum value of $z$ for each jet region determined
from equation (1) and equations (2) -(5), below. The ninth column is
the poloidal magnetic flux in the shell determined from the same
equations. The last three columns
define the electron energy power law from equation (1) that applies
throughout the shell.
\begin{table*}
\caption{Physical Parameters of the Jet Component Models}
{\footnotesize\begin{tabular}{cccccccccccc} \hline
Shell &  $Q$ & $N$ & B Field & $B^{\phi}$ & $B^{P}$ & $S$/KE & $S$/KE & $\Phi$&  $p$ & $\gamma_{\mathrm{min}}$ & $\gamma_{\mathrm{max}}$\\
          &  $10^{38}$ ergs/sec    & $\mathrm{cm}^{-3}$ &   & $10^{6}$ G & $10^{6}$ G & Base & Outer  & G - $\mathrm{cm}^{2}$&  &  & \\
 &  b  & a  &  a  & b  & b   & b    & b  & b  & a  & a   & a   \\

\hline
0   &  3.37 & $9.5 \times 10^{17}$ & Turbulent & 29.8 & 29.8 & 2.3 & $ 1.7 \times 10^{-2}$ & Net = 0 &   5.6 & 1.5 & 1000 \\
1   &  0.12  & $2.9 \times 10^{11}$ & MHD Jet & 6.9 & 3.3 & $ 4.8 \times 10^{4}$ & $2.1 \times 10^{4}$ & $1.4\times 10^{21}$ & 3.3 & 50 & 1000 \\
2   &  0.04  & $1.3 \times 10^{10}$ & MHD Jet & 2.6 & 2.3 & $ 3.4 \times 10^{5}$ & $2.6 \times 10^{5}$ & $8.4\times 10^{20}$&  3.3 & 50 & 1000 \\
3   &  0.18  & $2.7 \times 10^{9}$ & MHD Jet & 2.1 & 3.0 & $ 1.1 \times 10^{6}$ & $1.1 \times 10^{6}$ & $4.7\times 10^{21}$& 3.3 & 50 & 1000 \\
4   &  0.49  & $8.4 \times 10^{8}$ & MHD Jet & 1.1 & 2.9 & $ 1.0 \times 10^{6}$ & $1.0 \times 10^{6}$ & $2.0\times 10^{22}$&  3.3 & 50 & 1000 \\
\end{tabular}}
\caption{Table notes: a. input parameter for model fitted to SED
data, b. derived quantity from parameters of the jet model}

\end{table*}

\par The innermost shell labeled "0" (the core) is initiated in the
innermost part of the accretion disk. In the model, it resides in
the ergospheric portion of the disk inside of 1.7 M ($ M = 2\times
10^6 $ cm in geometrized units for the 14 $M_{\odot}$ black hole
mass inferred by \cite{gre01}). Thus, one requires a rapidly
spinning black hole and a black hole spin of $a/M \geq 0.99$ is
necessary to provide sufficient surface area in the disk. If a
slowly spinning black hole is used in the model, the volume of the base
of the jet increases dramatically and so does $Q$ (which is highly
dependent on the total amount of energized particles and magnetic
field - large jet volumes create large total plasma energy in the
models). The magnetic field is turbulent and it is assumed to be
advected with the jet as in \cite{bla79}. The jet begins near
equipartition as noted in Table 2. From equation 1, the value of $
m_{\mathrm{base}}=2.64 $ means that magnetic energy is dissipated
rapidly in the plasma near the base of the jet. The lost energy is
the X-ray emission forming the high energy power law. The core is
the most powerful portion of the jet and most of the energy is lost
to X-rays by inverse Compton cooling. The X-ray spectrum is computed
per the methods of \cite{tuc75,ghi96,sam97}. The X-ray power law
spectral index, $(p-1)/2$, fixes $p =5.6$. This is a very steep value
that is consistent with the assumption of intense Compton
cooling. The disk emission is fit by a simple blackbody (instead of
a parameterized accretion disk model) with $T=1.5 \times
10^{7}\,^{\circ}$ K and a surface area on each disk face of $2.4
\times 10^{14}\mathrm{cm}^{2}$ (the surface area of a thin Keplerian
disk for $r \leq 5M$ and $a/M=0.99$). The spectral fit has $\approx
4.0 \times 10^{38}$ ergs/sec in thermal disk luminosity and $\approx
2.5 \times 10^{38}$ ergs/sec in jet luminosity (note the total
matches that inferred by \cite{fuc04} when the flux overestimate
from RXTE/PCA of 16\% that they note is taken into account). The
large EUV peak in the synchrotron spectrum in Figure 2 is from the
same electrons that produce the EIC. The outer portion of the core
jet at large z is inertially dominated and is a weak synchrotron
source.
\par Shells 1 to 4 are Poynting flux dominated jets with a field line angular velocity, $\Omega_{F}$, equal to the
relativistic Keplerian velocity, $\Omega_{\mathrm{Kep}}$, at the
foot point \cite{lig75}. The field is organized and satisfies the
perfect MHD condition similar to the electrodynamic models of
\cite{bla76,lov76},
\begin{equation}
B^{\phi}= \frac{\rho B^{P}( d\phi/ dt - \Omega_{F})}{v^{P}}\;.
\end{equation}
The jets have an artificial simplification that the poloidal bulk
velocity is constant, $v^{P}=0.4c$. From equations (1) and (2), at
large $z$, the toroidal magnetic field, $B^{\phi}$, is much larger
than the poloidal magnetic field, $B^{P}$, and the particle angular
velocity, $d \phi/ dt$, is small compared to $\Omega_{F}$ (bounded velocity and angular
momentum). Perfect MHD requires that the total poloidal magnetic
flux is conserved in each shell,
\begin{equation}
\Phi = \int{B^{p} dA_{\perp}},
\end{equation}
where $dA_{\perp}$ is the cross-sectional area in the shell.
Equations (1) - (3) combine to give a relationship between $B^{P}$
at the base of the jet and $B^{\phi}$ at the start of the outer jet,
\begin{equation}
B^{P}_{\mathrm{base}} \approx
\left(\frac{v^{P}B^{\phi}_{\mathrm{outer}}}{\rho_{\mathrm{base}}\Omega_{\mathrm{Kep}}}\right)\left(\frac{z_{0}({\mathrm{outer}})}{z_{0}({\mathrm{base}})}\right)^{0.5}\;.
\end{equation}
This relationship allows one to compute (as in Table 2) $\Phi$ and
the poloidal Poynting flux, $S^{P}$, in both regions of the jet,
where
\begin{equation}
S^{P}=-\frac{1}{4\pi}\int{\Omega_{F} \rho B^{\phi}B^{p}
dA_{\perp}}\;.
\end{equation}
\par The base of the shell 1 jet produces the bulk of
the synchrotron IR emission. The jet emission in $K_{s}$ band was
estimated to be about 40\% of the total luminosity (the bulk being
from the donor star) in \cite{fuc03} consistent with the fit in
Figure 2. The value of $p =3.3$ for the shells is chosen to
reproduce the noticeable hardening of the X-ray spectrum above
$10^{19}$ Hz seen in the INTEGRAL data in Figure 2 and discussed
above. More than one half of this excess over the steep core
powerlaw arises from EIC at the base of the shell 1 jet. The other
shells provide the remainder of the hard EIC excess and the inner
part of the brightness profile of the jet in Figure 3. Shell 4 is
very energetic because the volume is huge - it has a very wide base.
Shells 2 and 3 are required to interpolate the brightness
distribution in Figure 3 between the "radio core" and base of the
observed radio jet produced by the plasma in shell 4 and the distant
regions of the brightness distribution that is emitted from plasma
in shell 1.
\par In summary, if one assumes that the hard X-ray power law is produced in the jet
(as in the fiducial model described in the last two sections) then the results in Table 2 imply that the jet power is
$Q =4.2\times 10^{38}$ ergs/sec. Assuming $a/M=0.99$, the mass flux in
the jet is $7.8 \times 10^{17}$ g/s.
\section{Parametric Sensitivity}
It is not claimed that this fiducial model is the unique jet solution that fits the data.
This raises the question of how the model depends on the parametric assumptions.
Although a formal parametric sensitivity analysis is
beyond the scope of this work, many trial models were produced and evaluated in the process of
arriving at the model presented in the previous sections. Here
are some of the observations that were made.
\begin{itemize}
\item \textbf{Number of Shells} The integrated kinematic quantities such as the magnetic, energy and mass fluxes
are not significantly affected by the number of shells. The more
shells, the smoother the brightness distribution and the better the
fit is in Figure 3. For example, if 4 or 6 shells were used instead
of 5, the SED can be reproduced. However, the fit in Figure 3, would
be considerably worse for 4 shells and somewhat better for 6 shells.
The luminosity in each shell decays on scales that are comparable to
the distance, $z$, that the peak of the luminosity is from the
accretion disk. For example, the 15 GHz luminosity is produced at
smaller $z$ than the 8.3 GHz luminosity in each shell. Thus, the
surface brightness dies off with a shorter scale length at 15 GHz
than at 8.3 GHz. Compare the right hand frame of Figure 3, to the
left hand frame - the 15 GHz fit is worse in the sense that there are 
large curved sections between actual data points and a pronounced 
inflection point neither of which exist within the sparsely sampled data 
(there are only 6 contour levels in 15 GHz image in \citet{fuc04}). This is manifested as the
almost exponential decay near 10 AU, followed by a sharp peak
at $\approx$ 25 AU. The more shells, the less pronounced the
alternating regions of exponential dips and sharp peaks become. With
4 shells, the 8.3 GHz plot would have similar large deviations from
the brightness profile as are seen in the existing 15 GHz plot in
Figure 3.
\item \textbf{Energy Flux Estimate} Given the assumptions in the model (in order of importance:
the hard X-ray luminosity is from the jet, the hard X-ray excess
above $10^{19}$ GHz is real and from the jet, and the IR excess
above the stellar contribution is from the jet), the model is very
close to a minimum energy flux. This is obvious since the energy
flux is only 1.5 times the radiated luminosity and the jet power
must exceed the broadband luminosity. The energy flux grows very
rapidly, if the jet is assumed to originate at large cylindrical
radius. The volume of plasma and fields increases rapidly and the
energy flux quickly rises above $10^{39}$ ergs/sec as the
cylindrical radii of the base of the shells increase by a factor of
$\, \GA \, 2$.
\item \textbf{Number Density} The number of energetic particles is fixed by the location of the base of the jet, the thermal X-ray flux
from the accretion disk and the constraints on the observed external
Compton luminosity. The particle flux and energy flux are
concentrated at small radius in the core. Driving most of the energy
and mass flux from the inner region of the disk is favored on
physical grounds. It co-locates the most energetic shell and largest
mass flux at the largest potential source of power to drive the jet
- the most energetic part of the accretion flow (fastest rotation
rate and highest temperature). Moving the core to the smallest
possible cylindrical radius also moves the other shells inward to
smaller radius and larger Keplerian velocity, thereby decreasing the
total volume and magnetic flux required to drive the flow, by
equation (5). This minimizes the energy flux in the outer shells. In
this decomposition, the mass flux in the outer shells is constrained
by the magnitude of the hard X-ray excess above $10^{19}$ Hz (EIC
from shells 1 -4) discussed in section 2. The large radio fluxes
from the outer shells combined with a modest EIC X-ray component
result in the magnetic dominance for the outer shells depicted in
columns 7 and 8 of Table 2. This analysis does not preclude the
possibility of a different transverse distribution of mass flux in a
jet that can reproduce the data, however the energy flux is likely
to be larger.
\item \textbf{Magnetic Field Strength} The magnetic field strength, $B$ is set by the
size of the base of the jet, the particle number density and the
synchrotron portion of the SED. The magnetic field must also be of
sufficient strength to be the ultimate energy reservoir that
supports the bulk of radiation losses as discussed in section 3.
\item \textbf{Electron Energy Spectrum} The electron spectrum is fixed by the X-ray power law spectrum as mentioned in section 2.
The only parameter that is not affected by the data is
$\Gamma_{max}$. Since the energy distributions are steep, there are
very few high energy electrons. A value of $\Gamma_{max} = 1000 $
was chosen arbitrarily. Any value larger than 700 will show
negligible differences in the results.
\item \textbf{Axial Dimensions} The value of $z_{0}(\mathrm{base})$ is set by the physical dimensions of the accretion/disk
black hole system. The quantities defined at $z_{0}(\mathrm{base})$
form the initial data for the initial surface of the entire jet. By
contrast, $z_{0}(\mathrm{outer})$ is set by the SED fit. For
example, large values of $z_{0}(\mathrm{outer})$ over-produce the
high frequency synchrotron emission. By definition
$z_{0}(\mathrm{outer})\equiv z_{\mathrm{max}}(\mathrm{base}) $. The
solution is insensitive to $z_{\mathrm{max}}(\mathrm{outer}) = 5
\times 10^{15}$ cm, since the jet emissivity is very low in this region.
\end{itemize}
\section{Alternative Models}
The shell decomposition in Table 2 is very useful for exploring variants of the
jet model. The most obvious variant is to assume that the disk corona and
not the jet is the source of the X-ray powerlaw. By lowering the mass
density and magnetic field density significantly (say, a factor of 100 for the sake of argument)
in the core, one can simply ignore the energy flux and
luminosity contribution to the composite jet. Then summing the remaining components in Table 2,
\begin{equation}
Q \approx 8.3\times 10^{37} \mathrm{ergs/sec\;,\, for\, a\,
coronal\, X-ray\, source,}
\end{equation}
and the jet plasma still reproduces the SED (less the coronal X-rays) and
the radio brightness profile in Figures 2 and 3, respectively.
\par There is uncertainty in the non-relativistic jet velocity as noted in Section 1.
Thus, it desirable to know the sensitivity of the jet power to
different values of jet poloidal velocity, $Q(v^{P})$. The shell
model lends itself to estimating $Q$ for different values of
$v^{P}$. First decompose $Q$ into 3 discrete pieces. Then note from
from equations (2) and (5), that $S^{P}$ in the outer jet has a
simple scaling with $v^{P}$, $S_{\mathrm{outer}} \sim
v^{P}(B^{\phi})^{2}$. Since $Q$ is a sum of radiation losses, the
mechanical kinetic energy flux in the outer jet, $KE$, and
$S_{\mathrm{outer}}$, one can use the results in Table 2 to
parameterize $Q$ as a function of $v^{P}$ when the X-ray power law
is from the jet. Implementing the Doppler enhancement factors from
\citet{lig75},
\begin{eqnarray} && Q(v^{P}) \approx
\left[\frac{\delta(\beta=0.4,\; \theta =70^{\circ})^{4}+
\delta(\beta=-0.4,\; \theta =70^{\circ})^{4}}{\delta(\beta,\; \theta
=70^{\circ})^{4}+\delta(-\beta,\; \theta =70^{\circ})^{4}}\right]\nonumber \\
&&\times (2.5\times 10^{38}) \nonumber \\
&& +\left[\frac{\delta(\beta=-0.4,\; \theta
=0^{\circ})}{\delta(-\beta,\; \theta =0^{\circ})}\right]^{4}
(\Gamma- 1)(v^{P}/0.436 c)(8.5 \times
10^{37})\nonumber \\
&& + (v^{P}/0.4c)(8.3 \times 10^{37}) \,
\mathrm{ergs/sec},
\end{eqnarray}
where $\delta \equiv \delta(\beta, \, \theta)$, the Doppler factor that
was introduced in Section 1, is a two parameter function of $\beta$ and $\theta$.
In this notation, $\delta(\beta=0.4,\; \theta =70^{\circ})$ means to evaluate
the Doppler factor at $\beta=0.4$ and $\theta =70^{\circ}$. The jet
poloidal velocity is $v^{P} \equiv c \beta$.
\par The Doppler correction on the first term in equation (7), means that larger $\delta$
factors require smaller intrinsic radiation losses to produce the
observed X-ray powerlaw SED in Figure 2. The Doppler correction on
the second term results from the redshifting of the disk thermal
emission in the rest frame of the plasma at the base of the jet. A
slow jet sees a stronger seed field and requires fewer hot particles
to reproduce the EIC SED and therefore has a smaller bulk kinetic
energy flux. The last term is the electromagnetic energy flux in the outer
jet. As noted in section 1, this formula will not hold for large
$\beta$ since SSC contributions become important.
\par Similarly, if the disk corona is the source of the X-ray powerlaw, one
can use the results in Table 2 and  equation (6) to parameterize the jet power as a function of
jet velocity as
\begin{eqnarray}
&& Q(v^{P}) \approx [v^{P}/0.4c][8.3 \times 10^{37}]\,
\mathrm{ergs/sec}\;.
\end{eqnarray}
\section{Discussion} The crude model presented here has demonstrated a few facts about the compact jet
in GRS 1915+105 in the "high plateau state":
\begin{itemize}
\item It is possible to model the jet in a manner analogous to AGN
\item The jet is powered during a high accretion state with a thermal luminosity of
$\approx 4\times 10^{38}$ ergs/sec, corresponding to an Eddington rate of $\approx$ 23\%.
\item The jet brightness profile requires a highly stratified jet in the transverse direction
\item Relativistic jet bulk velocity is not favored based on the huge implied
energy flux, $> 3.5 \times 10^{40}$ ergs/sec \textbf{if} the jet is the source of the X-ray power law.
\end{itemize}
\par An interesting consequence of the model was that a rapidly spinning black hole was preferred on energetic grounds.
This was not to tap black hole spin energy, but to use the surface area and high angular velocity of the plasma
in the ergosphere to power the jet core. It is interesting to note that the 3-D, perfect MHD, general relativistic
numerical simulations in \citet{haw06} find very high efficiency "funnel wall jets" emanating from this
region of the ergosphere of high spin black holes. The outflow is somewhat similar to what is described
in the core shell in this treatment,
near equipartition with turbulent disordered magnetic flux and an
outflow velocity of $ \sim $ 0.3c. The efficiency is sufficient to power the
core if $a/M >0.95$. They argue that the power source is pressure
gradients driven by the disk which seems energetically reasonable
in GRS 1915+105 considering the large thermal luminosity noted above.
\par Even though there is some numerical evidence to support the notion of a near-relativistic jet
core (where most of the energy transport occurs), it is not clear if
there is any analog of the compact jet in the AGN population.
In particular, are there near-relativistic outflows that are highly stratified
in the transverse direction that carry a mass flux comparable to the
accretion flux? It is also not obvious if there is any structure that can be construed as an analog of
a non-relativistic high luminosity, optically thick, radio jet in AGN.
If there is scale invariance in black hole
accretion systems then there should be an analog of the compact jet
in some AGN. The only
obvious way to reconcile this difference is to consider the
possibility that an AGN radio jet analogy is not appropriate and the
analog of the compact jet in AGN takes a different guise due to the
lower accretion disk temperature. The AGN feature that possesses the
majority of the characteristics of the jet model of GRS 1915+105 are the near-relativistic outflows
that are seen in X-ray absorption. A few radio quiet quasars have
been observed to have out-flowing, X-ray absorbing winds with
velocities between 0.1 c and 0.6 c \citep{cha03,gof11,pon03,sae09}.
The outflow in APM08279+5255 is the most energetic with a mass flux
similar to the accretion mass flux and a kinetic luminosity
comparable to the thermal luminosity of the accretion flow as in GRS
1915+105 \citep{sae09}. The fastest known out-flowing, X-ray
absorbing winds in radio loud AGN have $v < 0.15c$ \citep{tom10}.
\par In summary, due to the relatively low jet velocity and the large transverse stratification,
perhaps the compact jet in black hole binaries is best described as
a bipolar near-relativistic wind. In this analogy, it is not obvious
why the Galactic black hole version of the wind is luminous in the
radio band and the AGN version is not. 
\par It has been argued in \citet{kor06}
that there may exist a compact, optically thick jet in low-luminosity AGN.
This may or may not be applicable to the study here in that the thermal luminosity was 23\% of the Eddingtom limit
in these observations of GRS 1915+105. In any event, it is worthwhile to study jets from low luminosity
AGN with VLBI to look for evidence of nonrelativistic optically thick winds or jets. 
\par The implication of these jet models to other Galactic black holes can only be speculated upon because
other sources do not have the wealth of high resolution radio
images. Furthermore, GRS 1915+105 is a very strong X-ray source, so
there are many excellent studies of the radio jet/ X-ray connection.
Consider, the results of \citet{rus11,rod08} that the existence of a
hard power-law component is tantamount to the existence of a jet.
Within the framework of the assumption that inverse Compton emission
in the jet is the dominant source of the power law component, the
jet power can be estimated from the model presented here as
approximately 1.5 times the hard X-ray power law luminosity. This
estimate can be applied to other Galactic black holes in the hard
state. However, it would be much more accurate to constrain the
energy flux derived from the basic jet model on a case by case basis
with deep radio images and broadband simultaneous spectroscopy,
which are not readily available for other Galactic black holes. There
are estimates of the compact jet energy flux based on less direct methods
for other black hole candidates such as Cygnus X-1. For example, the emission from
a nebular ring 5 pc from Cygnus X-1 has been used to estimate
the energy flux required to energize the structure. This in turn is related to the compact jet power
on much smaller scales by assuming an invisible flow of energy to large scales.
This leads to estimates of the kinetic power of the compact jet of $10^{36}$ to $10^{37}$ ergs/sec \citep{gal05,rus07}.
This estimate is 0.3 to 1.0 of the total radiative luminosity, which is similar to
the jet power of approximately 0.6 times of the total radiative luminosity that was found for
the compact jet in GRS 1915+105 here.
\par Another consequence of the model is that a relativistic outflow
velocity in compact jets is not favored based on energetic concerns.
It is interesting to compare this to less direct estimates of the
compact jet speed in GRS 1915+105 and other Galactic black hole
candidates. Using the very basic jet model of \citet{bla79}, it is
argued in \citet{cas10} that $\Gamma > 2$ in GX 339-4, which was
have ruled out for GRS 1915+105 in the epoch considered here. A very
general argument was presented in \citet{hei04} that was based on
total X-ray and radio fluxes from a one zone model (that happens to
be cylindrical in shape). They found a loose set of constraints that
favored $\Gamma > 2$ in Galactic X-ray binaries, but was also
consistent with $\beta \approx 0.5$. Contrary to these works
\citet{gal03} used Monte Carlo simulations to argue that $\Gamma \,
\LA \, 2$ is preferred for Galactic X-ray binaries in the low/hard
state, marginally consistent with the energetic constraints of the
models presented here.

\section*{Acknowledgments} I am indebted to Jerome Rodriguez for analyzing the X-ray data
and discussing the implications with me. I would also like to thank
Marc Ribo for useful discussions regarding the radio data. I was
lucky to have a referee who offered many suggestions that improved
the presentation of the manuscript.

\end{document}